\documentclass[pra,showpacs,twocolumn,reprint,aps,]{revtex4-2}

\usepackage{amsfonts}
\usepackage{amsmath}
\usepackage{amssymb,amscd}
\usepackage{psfrag}
\usepackage{graphicx}
\usepackage{braket}
\usepackage[dvips]{epsfig}
\usepackage{epsfig}
\usepackage{subfigure}
\usepackage{color} 
\usepackage{amssymb}
\usepackage{lineno}

\begin{document}
\nolinenumbers
\title{Holevo Cram\'{e}r-Rao bound for multi-parameter estimation in nonlinear interferometers }
\author{Mengyao Zhou$^{1 }$}
\author{Hongmei Ma$^{2 }$}
\author{Liqing Chen$^{1 }$}
\author{Weiping Zhang$^{3,4,5,6}$}
\author{Chun-Hua Yuan$^{1}$}
\email{chyuan@phy.ecnu.edu.cn}
\date{\today }

\address{$^1$State Key Laboratory of Precision Spectroscopy, Quantum
	Institute for Light and Atoms, School of Physics and Electronic Science,
	East China Normal University, Shanghai 200062,China}
\address{$^2$School of Communication and Electronic Engineering, East China Normal University, Shanghai 200062,
China}
\address{$^3$School of Physics and Astronomy, and
	Tsung-Dao Lee Institute, Shanghai Jiao Tong University,  Shanghai 200240, China}
\address{$^4$shanghai Branch, Hefei National Laboratory, Shanghai 201315,
	China}
\address{$^5$Collaborative Innovation Center of Extreme Optics,
	Shanxi University, Taiyuan, Shanxi 030006, China}
\address{$^{6}$Shanghai
	Research center for Quantum Science, Shanghai 201315, China}

\begin{abstract}
Due to the potential of quantum advantage to surpass the standard quantum limit (SQL), nonlinear interferometers have garnered significant attention from researchers in the field of precision measurement. However, many practical applications require multiparameter estimation. In this work, we discuss the precision limit of multi-parameter estimation of pure Gaussian states based on nonlinear interferometers, and derive the Holevo Cram\'{e}r-Rao Bound (HCRB) for the case where both modes undergo displacement estimation. Furthermore, we compare our analytical results with the quantum Cramér-Rao Bound based on the symmetric logarithmic derivative (SLD-CRB), and with the result of the dual homodyne measurement. Through numerical analysis, we find that the HCRB equals the result of the dual homodyne measurement, whereas SLD-CRB is not saturable at small squeezed parameters. Therefore, this indicates that the HCRB is tight. Additionally, we provide intuitive analysis and visual representation of our numerical results in phase space.

\end{abstract}

\maketitle




\section{Introduction}

Precision measurement plays a crucial role in the research and development of physics, as it provides highly accurate measurement data. Interferometric measurement techniques, which rely on the interference phenomenon of waves to capture phase change information, are an indispensable tool for achieving high-precision measurements \citep{michelson1887,thompson2017}. This technique enables precise measurements of various physical quantities, such as angular velocity, magnetic field strength, and small distances. The phase measurement accuracy of traditional Mach-Zehnder interferometers (MZI) \citep{zehnder1891,mach1892} is limited by the standard quantum limit (SQL) \citep{braunstein1992,giovannetti2006}. In 1986, Yurke et al. proposed a new quantum interferometer scheme, the nonlinear SU(1,1) interferometer (SUI) \citep{yurke1986}. Compared to traditional MZI, the SUI can achieve the Heisenberg limit (HL) with fewer optical components under appropriate quantum state injection \citep{hudelist2014quantum,li2014phase,li2016phase}.

Quantum metrology, a highly prominent research field today, focuses on utilizing unique quantum-enhanced effects to achieve ultra-high precision measurements for specific parameters. The core mission of this field is to surpass the SQL in traditional measurements through quantum systems, enabling high-precision parameter estimation \citep{braunstein1994,giovannetti2004,giovannetti2011}. Within this framework, quantum parameter estimation forms the theoretical foundation of quantum metrology, encompassing both single-parameter and multi-parameter estimation processes. To date, there has been a wealth of theoretical and experimental research on single-parameter estimation, accumulating substantial insights and practical experience \citep{paris2009,albarelli2020,sahota2013,watanabe2010,ang2013,monras2010,lang2014,yuan2016,kok2017,jarzyna2012,takeoka2017,you2019}. In quantum parameter estimation, concepts such as the quantum Cramér-Rao bound based on the symmetric logarithmic derivative (SLD-CRB) \citep{braunstein1994} and quantum Fisher information (QFI) serve as crucial mathematical supports for single-parameter estimation research \citep{helstrom1969,holevo2011}.

However, it is worth noting that single-parameter estimation is often a simplified approach to complex real-world measurement models \citep{szczykulska2016}. In most practical applications, joint estimation of multiple parameters is essential. To address this, the QFI has been extended to the quantum Fisher information matrix (QFIM), and the QCRB has been generalized to a matrix inequality form to meet the needs of multi-parameter estimation. However, in quantum mechanics, the non-commutativity of certain observables leads to the fact that the optimal measurement for a single parameter in multi-parameter estimation may become incompatible. This characteristic gives rise to new precision limits for multi-parameter estimation, such as the Cramér-Rao bound based on the right logarithmic derivative (RLD-CRB) \citep{yuen1973,belavkin2004} and the Holevo Cramér-Rao bound (HCRB) \citep{holevo1973}.

Although the SLD-CRB and RLD-CRB provide valuable insights in some cases, they are generally not tight \citep{carollo2019,razavian2020,candeloro2021}, and the SLD-CRB and RLD-CRB do not fully reflect the interdependencies between parameters, which limits their application in multi-parameter estimation with nonlinear interferometers. In contrast, HCRB represents a stricter bound and can be achieved in the asymptotic limit through collective measurements of a large number of states\citep{massar1995,kahn2009,yamagata2013,yang2019}, and the difference between it and the SLD-CRB is bounded within a factor of two \citep{carollo2019,tsang2020}, and it is asymptotically achievable under certain conditions \citep{holevo2011,matsumoto2002}. Compared to the SLD-CRB, the HCRB can more accurately reflect the practical scenario of multi-parameter estimation in nonlinear interferometers and can overcome the limitations of the SLD-CRB, providing a more reliable lower bound for parameter estimation precision. Therefore, the HCRB has higher practical and theoretical value in multi-parameter quantum estimation.

However, using the HCRB to assess parameter estimation precision involves more complex calculations than the RLD-CRB and SLD-CRB, mainly because the optimization process of the HCRB requires handling nonlinear functions in the Hermitian matrix space \citep{albarelli2019}. Under certain conditions, such as for Gaussian states \citep{bradshaw2018,bradshaw2017tight}, pure states \citep{fujiwara1995,matsumoto2002}, estimation of qubit rotation \citep{conlon2021,conlon2023}, qubit systems \citep{suzuki2016,yung2024}, optical polarization \citep{jarzyna2022}, security issues in continuous-variable quantum communication protocols\citep{conlon2024}, and linear waveform estimation\citep{gardner2024},the corresponding analytical expressions for the HCRB can be derived. However, for multi-parameter estimation problems in nonlinear interferometers, no related research has yet been conducted. This gap motivates the present work.

In this paper, we extend the method of Ref.~\citep{bradshaw2017tight} to the nonlinear interferometer model \citep{kong2013phase}, considering the case where both modes experience unknown displacements. Based on the statistical model we establish, we derive the analytical expression for the HCRB and compute the results for the SLD-CRB. We evaluate and analyze the HCRB and SLD-CRB numerically, further verifying the effectiveness and accuracy of the HCRB. Additionally, we demonstrate through calculations that the  theoretical limit of the dual homodyne measurement are consistent with the HCRB results. This outcome indicates that the dual homodyne measurement represents the optimal achievable measurement, confirming that the HCRB is tight. Finally, to provide a more intuitive presentation and analysis of our research results, we use phase-space images for visualization. The results of this study not only provide new theoretical support and technical methods for multi-parameter estimation in nonlinear interferometers but also advance the field to a higher level.

The structure of our paper is as follows. In Section II, we delve into the theory of multi-parameter quantum local estimation, which provides crucial theoretical support for our subsequent research. In Section III, we describe in detail the measurement scheme for the displacement estimation of a two-mode Gaussian state based on a nonlinear interferometer, and  the analytical expression of the HCRB has been derived. In Section IV, we compare the HCRB with SLD-CRB and the results from the dual homodyne measurement. We also use the numerical analysis and phase-space images to intuitively analyze and discuss our results. In Section \ref{section:Conclusion}, we summarize the paper and provide an outlook on future research directions.

\begin{figure}[ht!]
\begin{center}
\centering{\includegraphics[scale=0.28,angle=0]{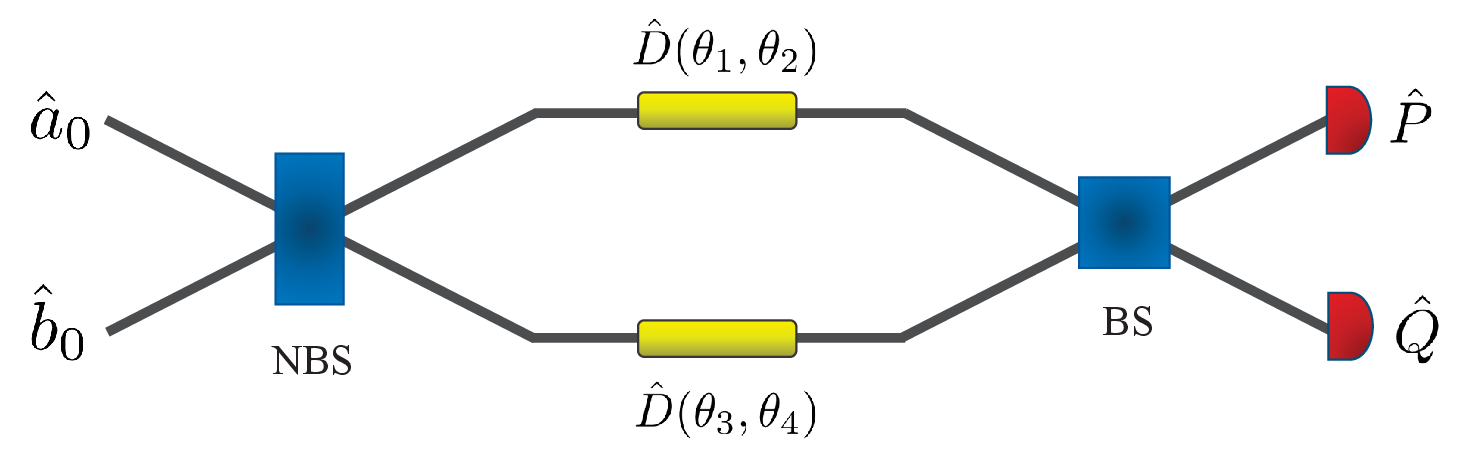}}
\end{center}
\caption{Schematic of the nonlinear interferometers for displacement estimation. The two modes of the light field go through parameter encoding for unknown displacements after passing through the nonlinear beam splitter (NBS), then they mix at the beam splitter (BS), and finally, the dual homodyne measurement is performed on the output states. $\hat{a}_{0}$ and $\hat{b}_{0}$ represent the two modes of the light field in the interferometer. }\label{fig1}
\end{figure}

\section{Multi-parameter Estimation Theory}
In this section, we give an overview of multi-parameter quantum estimation theory and its bounds including SLD-CRB, RLD-CRB and HCRB. 

Consider a set of quantum states $\rho_{\theta}$ that encodes $d$ parameters,
which can be represented as $\theta= (\theta_{1}, \theta_{2}, ..., \theta
_{d})$. Parameter estimation is performed by measuring the outcomes of
$\rho_{\theta}$ to estimate the value of $\theta$. In quantum mechanics,
measurements are described by a set of positive operator-valued measures
(POVM) $\hat{\Pi}= \{\hat{\Pi}_{\chi}\}$ \citep{nielsen2010}, where each measurement
outcome $\chi$ has a corresponding non-negative Hermitian operator $\hat{\Pi}_{\chi
}$, and the POVM elements sum to Identity: $\sum_{\chi} \hat{\Pi}_{\chi} = I$. The probability of obtaining the measurement result $\chi$ when the state
is $\rho_{\theta}$ is given by
$p_{\theta}(\chi) = \text{Tr}(\hat{\Pi}_{\chi} \rho_{\theta})$.
To obtain the estimated parameters based on the measurement result $\chi$, we
need an estimator $\hat{\theta}(\chi)$, which maps the measurement outcome
$\chi$ to an estimate of $\theta$. The accuracy of the estimator is typically
assessed using its mean squared error (MSE) matrix, denoted by $V_{\theta
}[\hat{\theta}]$, which is given by
\begin{equation}
V_{\theta}[\hat{\theta}] = \left[  \sum_{\chi} p_{\theta}(\chi) \left(
\hat{\theta}_{j}(\chi) - \theta_{j} \right)  \left(  \hat{\theta}_{k}(\chi) -
\theta_{k} \right)  \right]  _{jk}.
\end{equation}
An estimator is called a locally unbiased estimator if it satisfies $E\left[
\hat{\theta}(\chi) \right]  = \theta$ at the point $\theta$. It is referred to
simply as an unbiased estimator only if it is locally unbiased for all
corresponding $\theta= (\theta_{1}, \theta_{2}, ..., \theta_{d})$.

Assuming that $M$ measurements are independent, the corresponding MSE matrix
satisfies the Cram\'{e}r-Rao bound (CRB), which provides a lower bound for the
MSE matrix of a classical probability distribution $p_{\theta}(\chi)$:
\begin{equation}
V_{\theta}[\hat{\theta}] \geq\frac{1}{M} F^{-1},
\end{equation}
where $F$ is the classical Fisher Information (FI) matrix, defined as
\begin{equation}
\label{FI_Matrix}F_{jk} = \sum_{\chi} p_{\theta}(\chi) \left(  \partial_{j}
\log p_{\theta}(\chi) \right)  \left(  \partial_{k} \log p_{\theta}(\chi)
\right),
\end{equation}
where $\partial_{j} = \frac{\partial}{\partial\theta_{j}}$. This holds for any
fixed classical statistical model $p_{\theta}(\chi)$ \citep{cramer1946}. For
locally unbiased estimators, this bound is always achievable for any $M$. For
more realistic estimators, this bound is obtained in the limit of infinite
measurements $M$.

Due to this asymptotic achievability, especially when studying quantum
applications, locally unbiased estimators are often used
\citep{holevo2011,suzuki2020}, which corresponds to assuming that the
number of repetitions $M$ is sufficiently high to guarantee achievability.
Therefore, in the following analysis, we can ignore the factor $M$ from the
expressions. Thus, by choosing suitable effective estimators, we can achieve
at least the asymptotic CRB.

Moreover, quantum mechanics allows us to obtain a more general bound that
depends solely on the quantum statistical model $p_{\theta}(\chi)$ and is
independent of specific measurement strategies. Helstrom \citep{helstrom1967},
Yuen and Lax \citep{yuen1973}, and Belavkin \citep{belavkin2004} introduced the
SLD operator and the RLD operator, respectively.

The SLD operator is implicitly defined by the following Lyapunov equation:
\begin{equation}
\frac{\partial\rho_{\theta}}{\partial\theta_{i}}=\frac{\rho_{\theta}%
L_{i}+L_{i}\rho_{\theta}}{2}.
\end{equation}
For pure state models $\rho_{\theta}=|\psi_{\theta}\rangle\langle\psi_{\theta
}|$, the above equation can be easily solved, yielding a simple form for the
SLD operator:
\begin{equation}
L_{i}(\theta)=2\frac{\partial\rho_{\theta}}{\partial\theta_{i}}=2\left(
|\psi_{\theta}\rangle\langle\partial_{\theta_{i}}\psi_{\theta}|+|\partial
_{\theta_{i}}\psi_{\theta}\rangle\langle\psi_{\theta}|\right)
.\label{SLD_Define}%
\end{equation}
We can then define the entries of the QFIM
in terms of the SLD operator as follows:
\begin{equation}
Q_{ij}=\text{Tr}\left[  \rho_{\theta}\frac{L_{i}L_{j}+L_{j}L_{i}}{2}\right]
.\label{QFI_Matrix}%
\end{equation}
This can be used to derive a bound on the MSE, known as the SLD-CRB \citep{helstrom1969,helstrom1967}, denoted as
$C_{\theta}^{S}$:
\begin{equation}
\nu\geq C_{\theta}^{S}=\text{Tr}[Q_{ij}^{-1}].\label{SLD_CRB_Define}%
\end{equation}
Another QCRB we consider is defined based on the RLD operator,
which is given by:
\begin{equation}
\frac{\partial\rho_{\theta}}{\partial\theta_{i}}=\rho_{\theta}L_{i}.
\end{equation}
The entries of the QFIM in terms of the RLD operator are defined
as:
\begin{equation}
Q_{ij}^{R}=\text{Tr}\left[  \rho_{\theta}L_{i}L_{j}^{\dagger}\right]  .
\end{equation}
This leads to another bound on the MSE, known as the RLD-CRB \citep{yuen1973}, which we denote as $C_{\theta}^{R}$:
\begin{equation}
\nu\geq C_{\theta}^{R}=\text{Tr}\left\{  \text{Re}\left(  Q^{R}\right)
^{-1}\right\}  +\text{TrAbs}\left\{  \text{Im}\left(  Q^{R}\right)
^{-1}\right\}  .
\end{equation}
Here, $\text{TrAbs}\{A\}$ denotes the sum of the absolute values of the
eigenvalues of the matrix $A$. Although RLD-CRB and SLD-CRB are relatively
easy to compute \citep{paris2009,petz2011}, they are not always
achievable. 

Holevo derived another stricter bound for the MSE %
\citep{holevo2011,holevo2006noncommutative}, which we refer to as the HCRB. The HCRB is
defined through the following minimization: 
\begin{equation}
C_{\theta }^{H}:=\underset{\overrightarrow{X}\in \chi }{\min }h_{\theta }[%
\mathbf{\hat{X}}].  \label{HCRB_Min}
\end{equation}%
Here, $\mathbf{\hat{X}}=(\hat{X}_{1},\hat{X}_{2},...,\hat{X}_{d})$, and $%
\hat{X}_{j}$ is a Hermitian operator satisfying the following local unbiased
constraint conditions: 
\begin{align}
\text{Tr}(\rho _{\theta }\hat{X}_{j})& =0,  \label{HCRB_Req1} \\
\text{Tr}\left( \frac{\partial \rho _{\theta }}{\partial \theta _{j}}\hat{X}%
_{k}\right) & =\delta _{jk}.  \label{HCRB_Req2}
\end{align}%
Then, 
\begin{equation}
h_{\theta }[\mathbf{\hat{X}}]:=\text{Tr}\{\text{Re}Z_{\theta }[\mathbf{\hat{X%
}}]\}+\text{TrAbs}\{\text{Im}Z_{\theta }[\mathbf{\hat{X}}]\},
\end{equation}%
where $Z_{\theta }[\mathbf{\hat{X}}]$ is a $d\times d$ matrix defined as: 
\begin{equation}
Z_{\theta }[\mathbf{\hat{X}}]:=[\text{Tr}(\rho _{\theta }\hat{X}_{j}\hat{X}%
_{k})]_{j,k}.  \label{ztheta_Define}
\end{equation}%
For any Hermitian operator $\mathbf{\hat{X}}$ satisfying the local unbiased
constraint conditions, we have $h_{\theta }[\mathbf{\hat{X}}]\geq C_{\theta
}^{S}$ and $h_{\theta }[\mathbf{\hat{X}}]\geq C_{\theta }^{H}$. At the
minimum of $h_{\theta }$, we have $\nu \geq C_{\theta }^{H}$. Therefore, the
HCRB is always greater than or equal to the SLD-CRB and RLD-CRB.

\section{Precision bound-HCRB}

In this section, we calculate the HCRB for the nonlinear interferometer model shown in Fig.~\ref{fig1}. A coherent state  $\left\vert \alpha\right\rangle$ and a vacuum state $\left\vert 0
\right\rangle $ are injected. Thus, the input state takes the form of
\begin{equation}
\left\vert \alpha,0\right\rangle =\hat{D}(\alpha)\left\vert 0,0\right\rangle.
\end{equation}
where $\hat{D}(\alpha)$ is the displacement operator for the coherent state, defined
as
\begin{equation}
\hat{D}(\alpha)=\exp(\alpha\hat{a}_{0}^{\dagger}-\alpha^{\ast}\hat{a}_{0}).
\end{equation}

After passing through the nonlinear beam splitter (NBS), which corresponds
to the transformation induced by $\hat{U}_{NBS}=\exp (-\xi \hat{a}%
_{0}^{\dagger }\hat{b}_{0}^{\dagger }+\xi ^{\ast }\hat{a}_{0}%
\hat{b}_{0})$, where $\xi =\eta \alpha _{pump}^{2}t$, and $t$ is the
interaction time, the transformation of the annihilation operators is
\begin{equation}
\left( 
\begin{array}{c}
\hat{a}_{out} \\ 
\hat{b}_{out}%
\end{array}%
\right) =\left( 
\begin{array}{cc}
\cosh g & -e^{i\theta _{g}}\sinh g \\ 
-e^{-i\theta _{g}}\sinh g & \cosh g%
\end{array}%
\right) \left( 
\begin{array}{c}
\hat{a}_{in} \\ 
\hat{b}_{in}%
\end{array}%
\right) 
\end{equation}%
where $\xi =ge^{i\theta _{g}}$ \citep{gong2017intramode}, $g$ represents the strength of the nonlinear parametric process, and can also be referred to as the squeezed parameter. During the parameter encoding process after the NBS, the upper and lower
arms of the interferometer undergo unknown displacements of $\hat{D}(\theta
_{1},\theta _{2})=\exp (i\theta _{2}\hat{Q}_{1}-i\theta _{1}\hat{P}%
_{1})$ and $\hat{D}(\theta _{3},\theta _{4})=\exp (i\theta _{4}\hat{Q}%
_{2}-i\theta _{3}\hat{P}_{2})$ respectively. Here, $\hat{P}_{i}=(%
\hat{a}_{i}-\hat{a}_{i}^{\dagger })/2i$ and $\hat{Q}_{i}=(%
\hat{a}_{i}+\hat{a}_{i}^{\dagger })/2$ are a pair of orthogonal
operators. After this, the corresponding state transforms into $\left\vert
\psi \right\rangle $: 
\begin{equation}
\left\vert \psi \right\rangle =\hat{D}(\theta _{1},\theta _{2})\hat{D}(\theta
_{3},\theta _{4})\hat{U}_{NBS}\left\vert \alpha ,0\right\rangle .
\end{equation}

Finally, after passing through a 50:50 beam splitter (BS), we note that the
transformation associated with the BS is unitary. Therefore,
passing through the BS will not affect the ultimate precision limit
, and we can consider this process as part of the measurement scheme.
After the BS, the state changes to $\left\vert \psi\right\rangle
_{out}$:
\begin{align}
\left\vert \psi\right\rangle _{out}  & =\hat{U}_{BS}\hat{D}(\theta_{1},\theta_{2}%
)\hat{D}(\theta_{3},\theta_{4})\hat{U}_{NBS}\left\vert \alpha,0\right\rangle \nonumber\\
& =\hat{D}^{\prime}(\theta_{1},\theta_{2})\hat{D}^{\prime}(\theta_{3},\theta
_{4})\left\vert -\xi_{1a},\alpha^{\prime}\right\rangle\left\vert\xi_{1b},\beta^{\prime}\right\rangle,\label{Output_state}%
\end{align}
where $\left\vert -\xi _{1a},\alpha ^{\prime}\right\rangle $ and $\left\vert \xi _{1b},\beta ^{\prime }\right\rangle $ are
two single-mode squeezed coherent states, $\hat{D}^{\prime}(\theta_{1},\theta_{2})$ and $\hat{D}^{\prime}(\theta_{3}%
,\theta_{4})$ are the displacement operators after the BS
transformation:
\begin{align}
\hat{D}^{\prime}(\theta_{1},\theta_{2}) &  =\exp(i\theta_{2}\frac{1}{\sqrt{2}}%
\hat{Q}_{1}-i\theta_{1}\frac{1}{\sqrt{2}}\hat{P}_{1}\nonumber\\
&  +i\theta_{2}\frac{1}{\sqrt{2}}\hat{Q}_{2}-i\theta_{1}\frac{1}{\sqrt{2}}%
\hat{P}_{2}),
\end{align}%
\begin{align}
\hat{D}^{\prime}(\theta_{3},\theta_{4}) &  =\exp(i\theta_{4}\frac{1}{\sqrt{2}}%
\hat{Q}_{2}-i\theta_{3}\frac{1}{\sqrt{2}}\hat{P}_{2}\nonumber\\
&  -i\theta_{4}\frac{1}{\sqrt{2}}\hat{Q}_{1}+i\theta_{3}\frac{1}{\sqrt{2}}%
\hat{P}_{1}).
\end{align}

The displacement operators
$\hat{D}^{\prime}(\theta_{1},\theta_{2})$ and $\hat{D}^{\prime}(\theta_{3},\theta_{4})$
have the following partial derivatives with respect to $\theta_{1},\theta
_{2},\theta_{3},$ and $\theta_{4}$:
\begin{align}
\partial _{\theta _{1}}\hat{D}^{\prime }(\theta _{1},\theta _{2})& =[-\frac{i}{%
\sqrt{2}}\hat{P}_{1}-\frac{i}{\sqrt{2}}\hat{P}_{2}-\frac{i}{4}\theta _{2}]\hat{D}^{\prime
}(\theta _{1},\theta _{2}),  \notag \\
\partial _{\theta _{2}}\hat{D}^{\prime }(\theta _{1},\theta _{2})& =[\frac{i}{%
\sqrt{2}}\hat{Q}_{1}+\frac{i}{\sqrt{2}}\hat{Q}_{2}+\frac{i}{4}\theta _{1}]\hat{D}^{\prime
}(\theta _{1},\theta _{2}),  \notag \\
\partial _{\theta _{3}}\hat{D}^{\prime }(\theta _{3},\theta _{4})& =[-\frac{i}{%
\sqrt{2}}\hat{P}_{2}+\frac{i}{\sqrt{2}}\hat{P}_{1}-\frac{i}{4}\theta _{4}]\hat{D}^{\prime
}(\theta _{3},\theta _{4}),  \notag \\
\partial _{\theta _{4}}\hat{D}^{\prime }(\theta _{3},\theta _{4})& =[\frac{i}{%
\sqrt{2}}\hat{Q}_{2}-\frac{i}{\sqrt{2}}\hat{Q}_{1}+\frac{i}{4}\theta _{3}]\hat{D}^{\prime
}(\theta _{3},\theta _{4}).
\end{align}

To simplify the calculation, we compute the HCRB when $\theta$ is small. Thus,
we evaluate it at $\theta= 0$, because the bounds for all $\theta$ values are
the same. The reason we can do this is that the HCRB is asymptotically
achievable under adaptive measurement schemes, given a set of $n$ identical
states $\rho_{\theta}^{\otimes n}$, where $n \rightarrow\infty$. By using
$\sqrt{n}$ states for a small number of measurements, a rough estimate of
$\theta$ can be obtained, after which the remaining $n - \sqrt{n}$ states can
be replaced by $D(-\tilde{\theta})$, where $\tilde{\theta}$ is the
rough estimate of $\theta$, resulting in the state for small $\theta$.

Next, we compute $\left\vert \psi \right\rangle _{out}$ and the partial
derivatives of $\left\vert \psi \right\rangle _{out}$ with respect to $%
\theta _{1}$, $\theta _{2}$, $\theta _{3}$, and $\theta _{4}$ when $\theta
_{i}=0$: 
\begin{align}
\left\vert \psi _{0}\right\rangle & =\left\vert \psi \right\rangle
_{out}|_{\theta =0}=\left\vert -\xi _{1a},\alpha ^{\prime }\right\rangle
\left\vert \xi _{1b},\beta ^{\prime }\right\rangle ,  \notag \\
\left\vert \psi _{1}\right\rangle & =\partial _{\theta _{1}}\left\vert \psi
\right\rangle _{out}|_{\theta =0}  \notag \\
& =-\frac{i}{\sqrt{2}}[\hat{P}_{1}\left\vert -\xi _{1a},\alpha ^{\prime
}\right\rangle \left\vert \xi _{1b},\beta ^{\prime }\right\rangle
+\hat{P}_{2}\left\vert -\xi _{1a},\alpha ^{\prime }\right\rangle \left\vert \xi
_{1b},\beta ^{\prime }\right\rangle ],  \notag \\
\left\vert \psi _{2}\right\rangle & =\partial _{\theta _{2}}\left\vert \psi
\right\rangle _{out}|_{\theta =0}  \notag \\
& =\frac{i}{\sqrt{2}}[\hat{Q}_{1}\left\vert -\xi _{1a},\alpha ^{\prime
}\right\rangle \left\vert \xi _{1b},\beta ^{\prime }\right\rangle
+\hat{Q}_{2}\left\vert -\xi _{1a},\alpha ^{\prime }\right\rangle \left\vert \xi
_{1b},\beta ^{\prime }\right\rangle ],  \notag \\
\left\vert \psi _{3}\right\rangle & =\partial _{\theta _{3}}\left\vert \psi
\right\rangle _{out}|_{\theta =0}  \notag \\
& =-\frac{i}{\sqrt{2}}[\hat{P}_{2}\left\vert -\xi _{1a},\alpha ^{\prime
}\right\rangle \left\vert \xi _{1b},\beta ^{\prime }\right\rangle
-\hat{P}_{1}\left\vert -\xi _{1a},\alpha ^{\prime }\right\rangle \left\vert \xi
_{1b},\beta ^{\prime }\right\rangle ],  \notag \\
\left\vert \psi _{4}\right\rangle & =\partial _{\theta _{4}}\left\vert \psi
\right\rangle _{out}|_{\theta =0}  \notag \\
& =\frac{i}{\sqrt{2}}[\hat{Q}_{2}\left\vert -\xi _{1a},\alpha ^{\prime
}\right\rangle \left\vert \xi _{1b},\beta ^{\prime }\right\rangle
-\hat{Q}_{1}\left\vert -\xi _{1a},\alpha ^{\prime }\right\rangle \left\vert \xi
_{1b},\beta ^{\prime }\right\rangle ].
\end{align}

We are interested in the inner products of $|\psi_{0}\rangle$, $|\psi
_{1}\rangle$, $|\psi_{2}\rangle$, $|\psi_{3}\rangle$, and $|\psi_{4}\rangle$.
We set the coherent state corresponding to $\theta_{\alpha}=0$, and set
$\theta_{g}^{\prime}=\pi/2 $ for the NBS process, where $\theta
_{\alpha}=0$ means that $\alpha$ is a real number.
Under these circumstances, we present the results of the calculations of
five relevant inner products as follows:
\begin{align}
\langle\psi_{0}|\psi_{0}\rangle &  =1,\text{ }\langle\psi_{0}|\psi_{1}%
\rangle=-i|\alpha|\cosh(g),\\
\langle\psi_{0}|\psi_{2}\rangle &  =\langle\psi_{0}|\psi_{4}\rangle=0,\text{
}\langle\psi_{0}|\psi_{3}\rangle=-i|\alpha|\sinh(g).
\end{align}
We introduce a set of standard orthogonal basis states $\{|e_{0}\rangle, |e_{1}\rangle, |e_{2}\rangle\}$ that satisfy the corresponding inner product
conditions:
\begin{align}
\left\vert \psi_{0}\right\rangle  &  =|e_{0}\rangle,\nonumber\\
\left\vert \psi_{1}\right\rangle  &  =-i|\alpha|\cosh(g)|e_{0}\rangle
+\frac{i\sinh(g)}{2}|e_{1}\rangle-\frac{i\cosh(g)}{2}|e_{2}\rangle,\nonumber\\
\left\vert \psi_{2}\right\rangle  &  =\frac{-\sinh(g)}{2}|e_{1}\rangle
-\frac{\cosh(g)}{2}|e_{2}\rangle,\nonumber\\
\left\vert \psi_{3}\right\rangle  &  =-i|\alpha|\sinh(g)|e_{0}\rangle
+\frac{i\cosh(g)}{2}|e_{1}\rangle-\frac{i\sinh(g)}{2}|e_{2}\rangle,\nonumber\\
\left\vert \psi_{4}\right\rangle  &  =\frac{\cosh(g)}{2}|e_{1}\rangle
+\frac{\sinh(g)}{2}|e_{2}\rangle.
\end{align}

According to the local unbiased constraint condition
in Eqs. (\ref{HCRB_Req1}) and (\ref{HCRB_Req2}), they can be written as the
following two equations: 
\begin{eqnarray}
\langle e_{0}|\hat{X}_{1}|e_{0}\rangle  =\langle e_{0}|\hat{X}_{2}|e_{0}\rangle
=\langle e_{0}|\hat{X}_{3}|e_{0}\rangle =\langle e_{0}|\hat{X}_{4}|e_{0}\rangle =0, \nonumber\\
\end{eqnarray}%
and
\begin{eqnarray}
\langle \psi _{0}|\hat{X}_{k}|\psi _{j}\rangle +\langle \psi _{0}|\hat{X}_{k}|\psi
_{j}\rangle  =\delta _{jk},\text{ }(j,k=1,2,3,4). \nonumber\\ \label{constraint2}
\end{eqnarray}%
In Eq (\ref{constraint2}), each Hermitian operator $\hat{X}_{j}$ is subject to
four conditions, and there are a total of sixteen constraints.

To find the HCRB, we need to minimize the expression in Eq.~(\ref{HCRB_Min}).
Thus, we can set the components of $\hat{X}_{1},\hat{X}_{2},\hat{X}_{3}$, and $\hat{X}_{4}$ that do
not involve the above twenty constraints to their complex conjugates. We define
the components using their real and imaginary parts:
\begin{align}
\langle e_{0}|\hat{X}_{1}|e_{1}\rangle &  =t_{1}+ij_{1},\langle e_{0}|\hat{X}_{1}%
|e_{2}\rangle=s_{1}+ik_{1},\nonumber\\
\langle e_{0}|\hat{X}_{2}|e_{1}\rangle &  =t_{2}+ij_{2},\langle e_{0}|\hat{X}_{2}%
|e_{2}\rangle=s_{2}+ik_{2},\nonumber\\
\langle e_{0}|\hat{X}_{3}|e_{1}\rangle &  =t_{3}+ij_{3},\langle e_{0}|\hat{X}_{3}%
|e_{2}\rangle=s_{3}+ik_{3},\nonumber\\
\langle e_{0}|\hat{X}_{4}|e_{1}\rangle &  =t_{4}+ij_{4},\langle e_{0}|\hat{X}_{4}%
|e_{2}\rangle=s_{4}+ik_{4}.
\end{align}

The matrix in Eq.~(\ref{ztheta_Define}) is given by:
\begin{eqnarray}
&Z_{\theta}[\mathbf{\hat{X}}]=\nonumber\\
&\left(
\begin{array}
[c]{cccc}%
\text{tr}(\rho_{\theta}\hat{X}_{1}\hat{X}_{1}) & \text{tr}(\rho_{\theta}\hat{X}_{1}\hat{X}_{2}) &
\text{tr}(\rho_{\theta}\hat{X}_{1}\hat{X}_{3}) & \text{tr}(\rho_{\theta}\hat{X}_{1}\hat{X}_{4})\\
\text{tr}(\rho_{\theta}\hat{X}_{2}\hat{X}_{1}) & \text{tr}(\rho_{\theta}\hat{X}_{2}\hat{X}_{2}) &
\text{tr}(\rho_{\theta}\hat{X}_{2}\hat{X}_{3}) & \text{tr}(\rho_{\theta}\hat{X}_{2}\hat{X}_{4})\\
\text{tr}(\rho_{\theta}\hat{X}_{3}\hat{X}_{1}) & \text{tr}(\rho_{\theta}\hat{X}_{3}\hat{X}_{2}) &
\text{tr}(\rho_{\theta}\hat{X}_{3}\hat{X}_{3}) & \text{tr}(\rho_{\theta}\hat{X}_{3}\hat{X}_{4})\\
\text{tr}(\rho_{\theta}\hat{X}_{4}\hat{X}_{1}) & \text{tr}(\rho_{\theta}\hat{X}_{4}\hat{X}_{2}) &
\text{tr}(\rho_{\theta}\hat{X}_{4}\hat{X}_{3}) & \text{tr}(\rho_{\theta}\hat{X}_{4}\hat{X}_{4})
\end{array}
\right).\nonumber\\
\end{eqnarray}
Using the matrix $Z_{\theta}[\mathbf{\hat{X}}]$, we can compute the function
$h_{\theta}[\mathbf{\hat{X}}]$, and minimizing it leads to the HCRB result,
given as:
\begin{equation}
C^{H}=8e^{-2g}. \label{HCRB}%
\end{equation}
From the above expression, we find that the value of the HCRB decreases as the squeezed parameter $g$ increases.

\section{Discussion}\label{section:Discussion}

\begin{figure}[ht]
\begin{center}
\includegraphics[width=8cm]{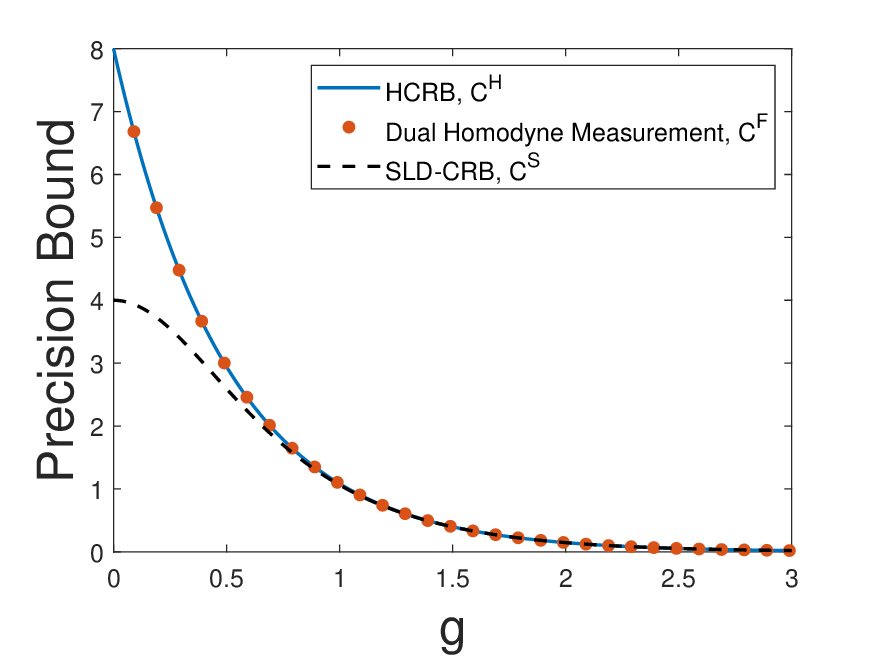}
\end{center}
\caption{Precision bound as a function of squeezed parameter $g$. The black dashed line represents the result of
SLD-CRB, the blue solid line shows the HCRB and the orange dotted line represents estimation precision
obtained using the dual homodyne measurement scheme. The diagram shows that the HCRB is consistent with the estimation precision achieved by the dual homodyne measurement scheme.}%
\label{fig2}%
\end{figure}

In this section, we make a comparison between the analytical results of HCRB and SLD-CRB, as well as the results from the dual homodyne measurement.

\subsection{ SLD-CRB}\label{subsection:Calculation of the Analytical Expression for SLD-CRB}

Based on the partial derivatives of the displacement operators with respect to
the parameters to be estimated, we can calculate the partial derivatives of
$\left\vert \psi\right\rangle _{out}$ with respect to the parameters.
Substituting these expressions into the definition of the SLD operator given
in Eq.~(\ref{SLD_Define}), we find the following SLD operators:
\begin{align}
L_{1}(\theta)  &  =2\left(  \left\vert \psi_{0}\right\rangle \langle\psi
_{1}|+\left\vert \psi_{1}\right\rangle \langle\psi_{0}|\right)  ,\nonumber\\
L_{2}(\theta)  &  =2\left(  \left\vert \psi_{0}\right\rangle \langle\psi
_{2}|+\left\vert \psi_{2}\right\rangle \langle\psi_{0}|\right)  ,\nonumber\\
L_{3}(\theta)  &  =2\left(  \left\vert \psi_{0}\right\rangle \langle\psi
_{3}|+\left\vert \psi_{3}\right\rangle \langle\psi_{0}|\right)  ,\nonumber\\
L_{4}(\theta)  &  =2\left(  \left\vert \psi_{0}\right\rangle \langle\psi
_{4}|+\left\vert \psi_{4}\right\rangle \langle\psi_{0}|\right)  .
\end{align}

Next, using the definition from Eq.~(\ref{QFI_Matrix}), we can derive the QFI matrix. After some calculations, we get the
following diagonal matrix:
\begin{equation}
Q^{S}=\left(
\begin{array}
[c]{cccc}%
\cosh(2g) & 0 & 0 & 0\\
0 & \cosh(2g) & 0 & 0\\
0 & 0 & \cosh(2g) & 0\\
0 & 0 & 0 & \cosh(2g)
\end{array}
\right)  .
\end{equation}
Finally, according to the definition given in Eq.~(\ref{SLD_CRB_Define}), we can
determine the corresponding SLD-CRB as follows:
\begin{equation}
C^{S}=\text{Tr}\left[  (Q^{S})^{-1}\right]  =\frac{4}{\cosh(2g)}.
\label{SLD_CRB}%
\end{equation}
From this expression, we see that under this statistical model, as the
squeezed parameter $g$ increases, the value of SLD-CRB decreases.
Therefore, in scenarios where only displacement estimation is involved,
squeezing is a valuable resource for improving the precision of displacement
estimation. The larger the squeezed parameter, the better the overall
estimation precision that can be achieved.

\subsection{Dual Homodyne Measurement}\label{subsection:Dual Homodyne Measurement}

By examining the equations corresponding to the statistical model in
Eq.~(\ref{Output_state}), we find that both modes carry information about the
displacement. Thus, when the two modes pass through a BS and are
squeezed in orthogonal directions in phase space, we can optimally estimate
the corresponding encoded parameters by performing appropriate orthogonal
measurements on the output modes. The dual homodyne measurement scheme
requires homodyne measurement of the orthogonal components of the two modes
separately, so from Eq.~(\ref{FI_Matrix}), we can derive the corresponding FI matrix as follows:
\begin{equation}
F=\frac{1}{2}\left(
\begin{array}
[c]{cccc}%
e^{2g} & 0 & 0 & 0\\
0 & e^{2g} & 0 & 0\\
0 & 0 & e^{2g} & 0\\
0 & 0 & 0 & e^{2g}%
\end{array}
\right)  .
\end{equation}

Therefore, the estimated precision from the dual homodyne measurement is
\begin{equation}
C^{F} = \text{Tr}[F^{-1}] = 8 e^{-2g}.
\end{equation}
This result indicates that this bound is consistent with our previously
obtained HCRB result. Since the dual homodyne measurement scheme is the
optimal measurement scheme for our statistical model, we conclude that the
HCRB is indeed tight.

As shown in Fig.~\ref{fig2}, the results of the HCRB,
SLD-CRB, and the sum of the MSE for the dual homodyne measurement as a function of the squeezed parameter $g$ are shown. We can see that both the HCRB and SLD-CRB values decrease as the squeezed parameter $g$ increases, indicating that
squeezing is a useful resource for enhancing the precision of displacement
estimation. Since the dual homodyne measurement is the optimal measurement
strategy, we find that the value of HCRB is equal to that of the dual homodyne measurement, which demonstrates that HCRB is tight. Furthermore, the
HCRB value is greater than that of SLD-CRB when the squeezed parameter $g$
is relatively small, indicating that using HCRB for evaluation is superior to
SLD-CRB. As the squeezed parameter $g$ increases beyond a certain point,
the values of both bounds converge.

\begin{figure}[ht]
\begin{center}
\includegraphics[width=6cm]{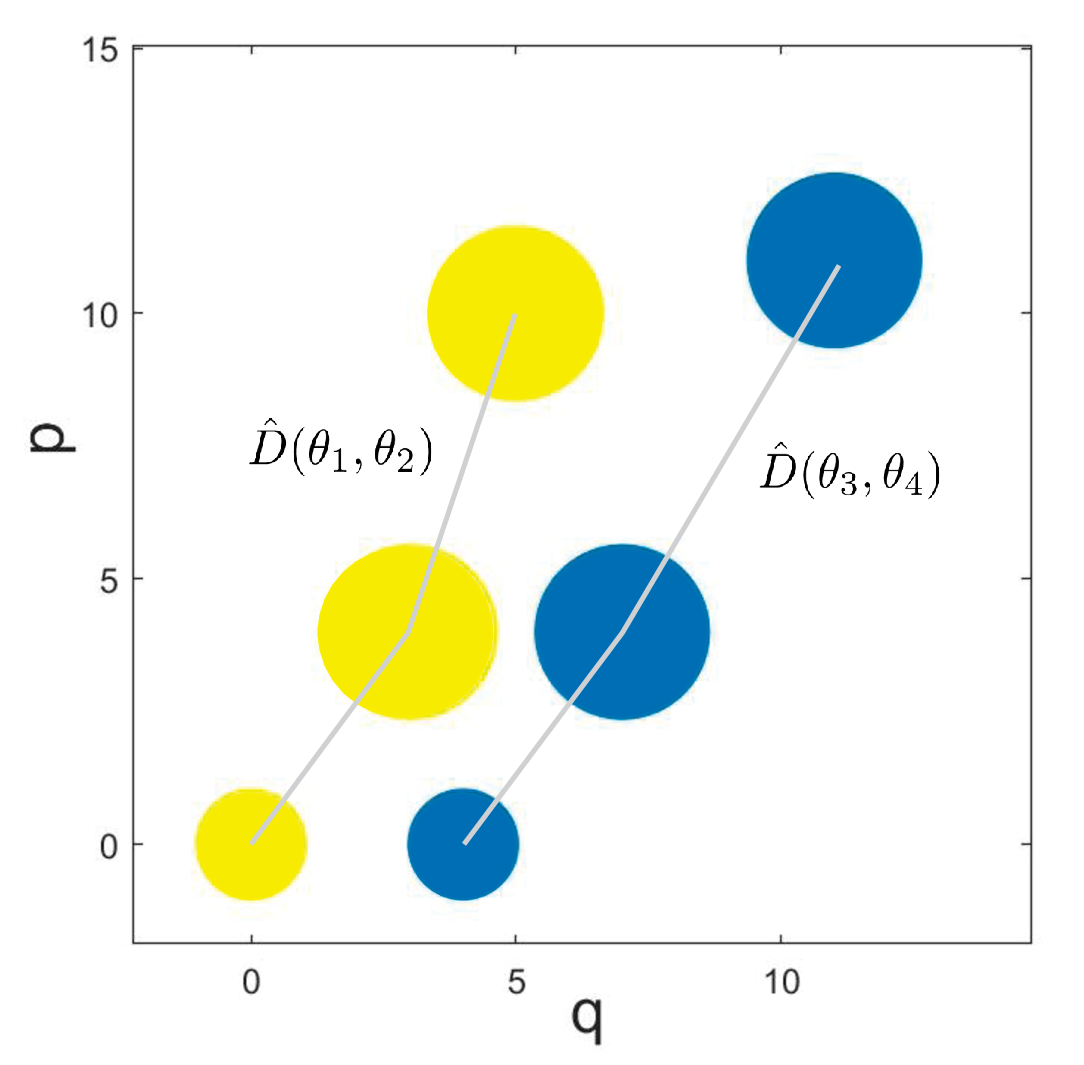}
\end{center}
\caption{Phase space evolution of the
two-mode light field after passing through the interferometer. The three
yellow circles on the left correspond to the transformations of mode $\hat
{a}_{0}$ in the interferometer, while the three blue circles on the right
correspond to the transformations of mode $\hat{b}_{0}$. The two smallest circles at the bottom represent the vacuum state and the coherent state. The two relatively larger circles in the middle and the two circles at the top respectively represent the phase space images of the light field after undergoing the NBS and the displacement encoding process.}%
\label{fig3}%
\end{figure}

Next, we provide an intuitive analysis and visualization of our numerical
results by illustrating the corresponding phase space diagrams of our
statistical model in different processes. The input states of the vacuum state and the coherent state can be represented by the yellow solid circle and the blue solid circle at the bottom, respectively, in the phase diagram of Fig.~\ref{fig3}. Since we set $\theta_{\alpha} = 0$ for the
coherent state, the corresponding phase diagram for the coherent state lies
along the axis where $p = 0$. After passing through the NBS, entangled two-mode squeezed light is generated. To facilitate our description in phase space, we consider the phase diagram of a single mode, where the individual mode distribution of a two-mode squeezed state follows a thermal distribution.The slightly larger solid yellow and
blue circles in the middle of Fig.~\ref{fig3} indicate that the fluctuations
of the orthogonal components of each mode have increased after passing through
the NBS, resulting in the larger circles in phase space. Finally, after
displacement encoding, the phase diagrams corresponding to the two modes have
unknown displacements in phase space. Based on the phase diagram, we can explain why the results we derived concerning HCRB and SLD-CRB do not include terms related to the number of photons. This is because we can treat the displacement throughout the process and the unknown encoded displacement as a unified entity. When discussing the corresponding limits of estimation accuracy, the fluctuations in the Orthogonal components remain unchanged.Consequently, when we are only interested in displacement
estimation, the resulting precision limits do not involve terms with photon numbers.

\section{Conclusion}\label{section:Conclusion}

In summary, we have provided a method for calculating the HCRB and analyzed
the estimation problem under the injection of coherent and vacuum states into
a nonlinear interferometer, followed by unknown displacement encoding. We have derived
the analytical expression for the HCRB corresponding to the statistical model,
and discussed its dependence on the squeezed parameter $g$. Our findings indicate that squeezing contributes to enhancing the HCRB, and that the HCRB results are superior to those of SLD-CRB, demonstrating that
HCRB is tight.

These results may offer methods for further improving measurement precision
and achieving practically reliable detection schemes. Currently, our derivation focuses on the parameter estimation problem for an ideal nonlinear interferometer with Gaussian pure state inputs. Our next step will involve extending this methodology to investigate related estimation issues concerning Gaussian mixed states in the presence of photon loss.

\section{ACKNOWLEDGMENTS}

This work is supported by the Shanghai Science and Technology Innovation Project No. 24LZ1400600, the Innovation Program for
Quantum Science and Technology 2021ZD0303200; the
National Natural Science Foundation of China Grants
No. 12274132, No. 11974111, No. 12234014,
No. 11654005, No. 11874152, and No. 91536114;
Shanghai Municipal Science and Technology Major
Project under Grant No. 2019SHZDZX01; Innovation
Program of Shanghai Municipal Education Commission
No. 202101070008E00099;
the National Key Research and
Development Program of China under Grant
No. 2016YFA0302001; and Fundamental Research
Funds for the Central Universities. W. Z. acknowledges
additional support from the Shanghai Talent Program.




\bibliography{test}
\nolinenumbers
\end{document}